\documentstyle[prb,aps,12pt,epsf,floats]{revtex}
\begin{document} 
   
\title{A new variational approach for the Holstein Molecular Crystal Model}
\author{V.~Cataudella, G.~De~Filippis and 
G.~Iadonisi} 
\address{$^\dagger$Dipartimento di Scienze Fisiche, 
Universit\`a di Napoli I-80125 Napoli, Italy}  
\date {\today} 
\maketitle 
\begin {abstract} 
A new variational technique is developed to investigate the polaronic 
features of the Holstein Molecular Crystal Model. 
It is based on a 
linear superposition of Bloch
states that describe 
large and small polaron wave functions. 
It is shown that this method  
provides a very good description of the regime  
characterized 
by intermediate 
values of the electron-phonon coupling constant (the so-called intermediate 
polaron) for any value 
of the 
adiabatic parameter $\omega_0/t$. 
The polaron ground state energy in 
one and two dimensions is calculated and  
successfully compared with the best 
estimates available providing a clear physical 
interpretation of the intermediate polaron. 
The band structure, the spectral weight of the ground state 
and the lattice displacement associated to the polaron 
are also calculated and discussed. 
The new method has 
the advantage to require a very little computational effort. 
  
\end {abstract} 

\newpage
\section {Introduction}

Recently a large amount of experimental results, 
ranging from infrared spectroscopy to 
transport properties involving the colossal magneto-resistance and high $T_c$ 
superconductivity, has pointed out the presence of polaronic carriers in doped 
cuprates and in the manganese oxide perovskites\cite{1,2}. 
In particular several experiments have shown that in doped perovskite 
manganites $La_{1-x}A_{x}MnO_{3}$ ($A=Sr, Ca$) there is a quite 
considerable 
coupling between the charge and the lattice degrees of freedom\cite{2}. 
The lattice 
distortions associated with the $Mn$ ions play an important role in  
determining the electronic and magnetic properties of these compounds which 
have become the focus of the scientific interest after the discovery of the 
colossal magneto-resistance phenomena. Recently it has been found that 
in $La_{0.75}Ca_{0.25}MnO_{3}$ the metallic phase is characterized by 
homogeneously distributed intermediate polarons while, above the transition 
from ferromagnetic metal to paramagnetic insulator, small and 
intermediate polarons coexist\cite{3}. Also the measurements of the $CuO$ 
distances in $La_{1.85}Sr_{0.15}CuO_{4}$ crystal have pointed out, 
below $100 K$, two 
conformations of the $CuO_{6}$ octahedra assigned 
to two different types of polarons\cite{Bianconi}.  

This large amount of experimental data has renewed the interest in studying 
the models of the electron-phonon coupled system and in particular the 
Holstein 
molecular crystal model that, for his relative simplicity, 
is the most considered model for the interaction of a single tight-binding 
electron coupled to an optical local phonon mode\cite{4}. 

For the Holstein hamiltonian, beside the weak-coupling perturbative 
theory\cite{5} 
an analytical approach is known for the strong 
coupling limit in the nonadiabatic regime (small polaron)\cite{6}. 
It is based on the Lang-Firsov 
canonical transformation and on 
expansion in powers of $1/\lambda$ where $\lambda=E_p/zt$ is the 
dimensionless coupling constant, $E_p$, $z$ and $t$  being, respectively, 
the small polaron binding energy, the coordination lattice number and 
the bare 
effective hopping integral ($\lambda$ represents the ratio between the 
small polaron binding energy and the energy gain of an itinerant electron on 
a rigid lattice). It is well known that 
both these analytical techniques fail to 
describe the region, of greatest physical interest, characterized by 
intermediate couplings and by electronic and phononic energy scales not well 
separated. This regime has been analyzed in several works based on Monte 
Carlo simulations\cite{7,korn}, 
numerical exact diagonalization of small clusters\cite{8}, 
dynamical mean field theory\cite{9}, density matrix renormalization 
group\cite{10} and variational approaches\cite{11,17}. 
The general conclusion is that the ground state energy and the effective 
mass in the Holstein model are continuous functions of the electron-phonon 
coupling and that there is not phase transition in this one-body system
\cite{Lowen}. 
In particular when the interaction strength is greater than a critical value 
the ground state properties change significantly but without breaking the 
translational symmetry. 

Recently, results for the Holstein molecular crystal model 
have been presented by using the Global-Local variational 
method\cite{11}. 
The comparison of the data obtained in one space dimension 
with the known approaches has shown that these results 
seem to provide the best estimates of the polaron ground state energy. 
They are highly accurate over a 
wide range of the polaron parameter space, from the non-adiabatic to the 
adiabatic, from weak to strong coupling limit. Nevertheless a 
solution of the Global-Local variational method for any particular $k$ 
value ($k$ is the wave number of the polaron Bloch state) 
is obtained by minimizing with respect to a very large number 
of parameters, that depends on the number of lattice sites and 
that increases dramatically with increasing the number 
of space dimensions from one to three.   

The aim of this work is to study the Holstein polaron features including 
ground state energy, polaron energy band, shape of the lattice distortion 
induced by the electron-phonon interaction and spectral weight of the 
coherent polaron band within a new variational approach. 
It is based on two translationally invariant Bloch wave functions that provide 
a very good description of the two asymptotic regimes, the weak and strong 
coupling regimes. In this paper these wave functions are called 
large and small polaron. 
In the large polaron wave function the phonon distribution 
function takes into account the average effect of the correlation among 
the emission of successive virtual phonons by the electron and the spatial 
extension of the polaron is large compared with the lattice parameter of the 
crystal. In the small polaron wave function 
the lattice polarization is confined 
to a region of the order of the unit cell and the polaron radius becomes of 
the order of the lattice constant. In this case all momenta of the Brillouin 
zone contribute to the polaron wave function and in the phonon distribution 
function the effect of the electron recoil due to the emission of virtual 
phonons is negligible. 

A careful inspection of these two wave functions points out that, far away 
from the two asymptotic regimes, they are not orthogonal 
and that the off-diagonal matrix elements of the Holstein hamiltonian 
are not zero. It is then straightforward 
to determine variationally the polaron ground 
state energy by considering as trial state the linear superposition 
of the large and small polaron wave functions. 

The comparison of our results with the Monte Carlo\cite{korn} data and the 
ground state energies of the variational global local method\cite{11} 
shows that the proposed method provides a very good description 
of the polaron ground state energy for any value of the parameters 
of the Holstein model and confirms the existence of three 
regimes\cite{11}: 
the weak coupling regime, characterized by polaron masses lightly heavier 
than the free electron mass and by dimension of the lattice polarization 
large compared with the lattice parameter; the strong coupling regime where 
the well-known polaronic band collapse takes place and the intermediate 
regime that is characterized by 
the crossover between the small and large polaron solutions. 
This regime is, therefore, 
well described by a wave function that is a linear 
superposition of Bloch states that describe the small and large polaron. 

We stress that the new variational approach provides a clear description 
of the Holstein polaron features in any regime and involves, for any 
particular $k$ value, a very small number of variational parameters, that 
does not depend on the number of lattice sites. 

\section {A new variational wave function}
  
{\bf The model}. The Holstein molecular crystal model is described by the 
Hamiltonian\cite{4}: 
\begin{eqnarray} 
&&H=H_{el}+H_{ph}+H_{I}=-t\sum_{<i,j>} c^{\dagger}_{i}c_{j}
+\omega_0\sum_{\vec{q}}a^{\dagger}_{\vec{q}}a_{\vec{q}}
+\sum_{i,\vec{q}} 
c^{\dagger}_{i}c_{i}\left[M_{q}e^{i\vec{q}\cdot \vec{R}_{i}}a_{\vec{q}}
+h.c.\right]~.\nonumber \\
&&
\label{1r}
\end{eqnarray}

The units are such that $\hbar=1$. Since we will restrict ourselves to the 
single electron case we will not consider electron spin indices. 
The symbol $< >$ in the first term of the sum in Eq.(\ref{1r}) means that
the summation is to be carried out only when $i$ and $j$ are nearest
neighbours to each other. 

In the Eq.(\ref{1r}) $c^{\dagger}_{i}$ denotes the electron creator 
operator at site $i$, the position vector of which is indicated by 
$\vec{R}_{i}$, $a^{\dagger}_{\vec{q}}$ represents the creation operator 
for phonon with wave number $\vec{q}$, $t$ is the transfer integral between 
nearest neighbor sites, $\omega_0$ is the frequency of the optical local 
phonon mode and $M_{q}$ indicates the electron-phonon matrix 
element. In the Holstein model (short range electron-phonon interaction) 
$M_{q}$ assumes the form: 
\begin{equation}
M_{q}=\frac{g}{\sqrt{N}} \omega_0 .
\label{2r}
\end{equation}
Here $N$ is the number of 
lattice sites.

{\bf The small polaron} 

When the value of $g$ is sufficiently large the lattice 
polarization cannot follow the electronic oscillations and, therefore, 
depends only on 
the average charge distribution of the electron. The wave function of the 
system can be factorized into a product of 
normalized variational functions 
$|\varphi>$ and $|f>$ depending on the electron and phonon coordinates 
respectively\cite{pekar}: 
\begin{equation}
|\psi^{(s)}>=|\varphi>|f>
\label{5r}
\end{equation}   
where 
\begin{equation}
|\varphi>=\sum_{\vec{R}_m}c^{\dagger}_m |0>_{el} \phi(\vec{R}_m)
\label{6r}
\end{equation}
and $|f>$ has to be determined variationally.

In the Eq.(\ref{6r}) 
$|0>_{el}$ is the electron vacuum state and  $\phi(\vec{R}_m)$ are
variational parameters that satisfy the relation:    
\begin{equation}
\sum_{\vec{R}_m}|\phi(\vec{R}_m)|^2=1~.
\label{7r}
\end{equation}

The expectation value of the Hamiltonian (\ref{1r}) on the state 
(\ref{5r}) gives: 
\begin{eqnarray}
&&<\psi^{(s)}|H|\psi^{(s)}>=-t\sum_{\vec{R}_m,<\vec{\delta}>}\phi^*(\vec{R}_m)
\phi(\vec{R}_m-\vec{\delta})+<f|\sum_{\vec{q}}\left[
\omega_0a^{\dagger}_{\vec{q}}a_{\vec{q}}+\rho_{\vec{q}}a_{\vec{q}}
+\rho^*_{\vec{q}}a^{\dagger}_{\vec{q}}
\right]|f>\nonumber \\
&&
\label{8r}
\end{eqnarray} 
with 
\begin{equation}
\rho_{\vec{q}}=M_{q}\sum_{i}e^{i\vec{q}\cdot \vec{R}_i}|\phi(\vec{R}_i)
|^2~.
\label{9r}
\end{equation}
In the Eq.(\ref{8r}) the symbol $< >$ in the summation means 
that $\vec{\delta}$ runs only over the 
nearest neighbours. 

The variational problem with respect to $|f>$ leads to the following 
lowest energy phonon state: 
\begin{equation}
|f>=e^{\sum_{\vec{q}}\left[ \frac{\rho_{\vec{q}}}{\omega_0}a_{\vec{q}}
+h.c.\right]}|0>_{ph},
\label{10r}
\end{equation} 
where $|0>_{ph}$ is the phonon vacuum state, 
and to the following total energy:
\begin{equation}
E_0=-t\sum_{\vec{R}_m,<\vec{\delta}>}\phi^*(\vec{R}_m)\phi(\vec{R}_m-
\vec{\delta})-\sum_{\vec{q}}\frac{|\rho_{\vec{q}}|^2}{\omega_0}~.
\label{11r}
\end{equation}

Even if this self-trapped state provides a good approximation of the 
ground state energy of the small polaron it is evident that the true 
eigenstate of the electron-lattice coupled system has translational 
symmetry. 
We construct translationally invariant 
Bloch states  by taking a superposition of the 
localized states (\ref{5r}) centered on different lattice sites 
in the same manner in which 
one constructs a Bloch wave function from 
a linear combination of atomic orbitals. Then the trial wave function that 
accounts for the translational symmetry is given by (see also ref.17): 
\begin{equation}
|\psi^{(s)}_{\vec{k}}>=\frac{1}{\sqrt{N}}\sum_{\vec{R}_n}e^{i\vec{k}\cdot 
\vec{R}_n}|\psi^{(s)}_{\vec{k}}(\vec{R}_n)>
\label{12r}
\end{equation}
where 
\begin{equation}
|\psi^{(s)}_{\vec{k}}(\vec{R}_n)>
=\sum_{\vec{R}_m}c^{\dagger}_{m+n}|0>_{el}
\phi_{\vec{k}}(\vec{R}_m)
e^{\sum_{\vec{q}}\left[ f_{\vec{q}}(\vec{k})a_{\vec{q}}
e^{i\vec{q}\cdot \vec{R}_n}
+h.c.\right]}|0>_{ph}
\label{13r}
\end{equation} 
and 
\begin{equation}
f_{\vec{q}}(\vec{k})=\frac{\rho_{\vec{q}}(\vec{k})}{\omega_0}=
\frac{M_{q}}{\omega_0}\sum_{\vec{R}_m}|\phi_{\vec{k}}(\vec{R}_m)|^2
e^{i\vec{q}\cdot \vec{R}_m}~.
\label{14r}
\end{equation}

This wave function is a sum of coherent states in the phonon coordinates, 
one for any particular lattice site. Since in a coherent state the emission 
of phonons occurs through a number of independent processes it is evident 
that in this trial state there is not correlation among the 
emission of successive virtual phonons. 
This physical assumption, i.e. that on 
every lattice site virtual phonons 
are emitted independently, is well-founded when $g$ 
is sufficiently large but it is questionable for intermediate and small values 
of the electron-phonon interaction where the electron recoil kinetic 
energy plays an essential role.  
Moreover the wave function (\ref{12r}) does not contain states with real 
phonons. This indicates that the calculation of the polaron energy provides 
correct results only if the effective polaron band width $\Delta$ and the 
phonon energy $\omega_0$ satisfy the condition $\Delta< \omega_0$. 
As it is well known, 
both these approximations limit the validity of the wave function (\ref{12r}) 
to the strong coupling limit. 

The expectation value of the Hamiltonian (\ref{1r}) on the state 
(\ref{12r}) gives: 
\begin{eqnarray}
&&<\psi^{(s)}_{\vec{k}}|H_{el}|\psi^{(s)}_{\vec{k}}>=
-t\sum_{\vec{R}_n}e^{i\vec{k}\cdot \vec{R}_n}
e^{-\sum_{\vec{q}} |f_{\vec{q}}|^2 \left(1-e^{-i\vec{q}\cdot \vec{R}_n}
\right)} \sum_{\vec{R}_m,<\vec{\delta}>}\phi^*_{\vec{k}}(\vec{R}_m)
\phi_{\vec{k}}(\vec{R_m}-\vec{R}_n-\vec{\delta}) \nonumber\\
&&
\label{1e}
\end{eqnarray}
\begin{eqnarray}
&&<\psi^{(s)}_{\vec{k}}|H_{ph}+H_{I}|\psi^{(s)}_{\vec{k}}>=
\sum_{\vec{R}_n}e^{i\vec{k}\cdot \vec{R}_n}
e^{-\sum_{\vec{q}} |f_{\vec{q}}|^2 \left(1-e^{-i\vec{q}\cdot \vec{R}_n}
\right)} \sum_{\vec{R}_m}
\phi^*_{\vec{k}}(\vec{R}_m)
\phi_{\vec{k}}(\vec{R_m}-\vec{R}_n)
\nonumber \\ 
&&\sum_{\vec{q}}\left[
|f_{\vec{q}}|^2 \omega_0
e^{-i\vec{q}\cdot \vec{R}_n}
-f^*_{\vec{q}}M_{q} 
e^{i\vec{q}\cdot \left(\vec{R}_m-\vec{R}_n\right)}
-f_{\vec{q}}M^*_{q} 
e^{-i\vec{q}\cdot \vec{R}_m}\right]~.
\label{15r}
\end{eqnarray}
 
Moreover the calculation of the normalization factor 
$<\psi^{(s)}_{\vec{k}}|\psi^{(s)}_{\vec{k}}>$ gives: 
\begin{equation}
<\psi^{(s)}_{\vec{k}}|\psi^{(s)}_{\vec{k}}>=
\sum_{\vec{R}_n}e^{i\vec{k}\cdot \vec{R}_n}
e^{-\sum_{\vec{q}} |f_{\vec{q}}|^2 \left(1-e^{-i\vec{q}\cdot \vec{R}_n}
\right)} \sum_{\vec{R}_m}
\phi^*_{\vec{k}}(\vec{R}_m)
\phi_{\vec{k}}(\vec{R_m}-\vec{R}_n)~.
\label{16r}
\end{equation}

The next step is the determination of the variational parameters 
$\phi_{\vec{k}}(\vec{R}_n)$. We note that if one neglects the spatial 
broadening of the electronic wave function, i.e. 
$\phi_{\vec{k}}(\vec{R}_n)=\delta_{\vec{R}_n,0}$, 
the Lang Firsov approximation is recovered, i.e. the exact solution of the 
Holstein molecular crystal model for $\omega_0/t \rightarrow \infty$. 
When the value of the adiabatic parameter $\omega_0/t$ decreases 
it becomes necessary to go beyond this 
approximation. 
In this paper we assume: 
\begin{equation}
\phi_{\vec{k}}(\vec{R}_n)=\alpha_{\vec{k}} \delta_{\vec{R}_n,0}+
\beta_{\vec{k}} \delta_{\vec{R}_n,\vec{\delta}}+
\gamma_{\vec{k}}\delta_{\vec{R}_n,\vec{\zeta}}~.
\label{17r}
\end{equation}  
Here $\beta_{\vec{k}}$ and $\gamma_{\vec{k}}$ are two variational 
parameters, $\alpha_{\vec{k}}$ is determined in such a way the 
Eq.(\ref{7r}) is satisfied and 
$\vec{\delta}$ and $\vec{\zeta}$ indicate, respectively,  
the nearest and the next 
nearest neighbours. 

This choice of the parameters $\beta_{\vec{k}}$ and $\gamma_{\vec{k}}$ that 
takes into account the broadening of the electronic wave function in every 
lattice site to the nearest neighbours and to the next nearest neighbours 
allows to obtain a variational estimate of the ground state energy 
at $\vec{k}=0$ 
that is lower than the result of the second order of the 
perturbation theory\cite{6}: 
\begin{equation}
E^{(sc)}\simeq E_p \left(1+\frac{1}{2z\lambda^2}\right)
\label{18r}
\end{equation} 
where $E_p$ indicates the small polaron binding energy to the first order 
of the perturbation theory:
\begin{equation}
E_p=-\sum_{\vec{q}}\frac{|M_q|^2}{\omega_0}
\label{19r}
\end{equation}
and $z$ is the nearest neighbour number. 
In appendix the calculation of the polaron band 
within the ansatz (\ref{17r}) for the variational parameters 
$\phi_{\vec{k}}(\vec{R}_n)$ is reported. 

We end this section noting that this method can be systematically 
improved by adding further terms in 
Eq.(\ref{17r}). This allows to obtain 
better and better estimates of the polaron energy 
in the strong coupling limit.    

{\bf The large polaron}

It is well known that 
when the value of $g$ is small the picture is 
quite different. 
As the electron moves through the crystal it exerts weak 
forces upon the ions which respond and move. This resultant ionic 
polarization will, in turn, modify the motion of the electron. Then 
the particle must drag this polarization with it during its motion through 
the solid. This affects its effective mass that is weakly larger than 
that of a Bloch electron\cite{12}. In this weak-coupling regime is useful 
to adopt a 
variational approach similar to that of Lee, Low and Pines in the 
continuum approximation\cite{14}. 
A possible choice for the trial wave function is: 
\begin{equation}
|\psi^{(l)}_{\vec{k}}>=\frac{1}{\sqrt{N}}\sum_{\vec{R}_n}e^{i\vec{k}\cdot 
\vec{R}_n}|\psi^{(l)}_{\vec{k}}(\vec{R}_n)>
\label{20r} 
\end{equation}
where
\begin{equation}
|\psi^{(l)}_{\vec{k}}(\vec{R}_n)>=c^{\dagger}_{n}|0>_{el}
e^{\sum_{\vec{q}}\left[h_{\vec{q}}(\vec{k})a_{\vec{q}}
e^{i\vec{q}\cdot \vec{R}_n}
+h.c.\right]}
\left[|0>_{ph}+\sum_{\vec{q}}d^*_{\vec{q}}(\vec{k})
e^{-i\vec{q}\cdot \vec{R}_n}
a^{\dagger}_{\vec{q}}|0>_{ph}\right]
\label{21r}
\end{equation}
and 
\begin{equation}
h_{\vec{q}}(\vec{k})=
\frac{M_{q}}{\omega_0+E_b(\vec{q})-E_b(\vec{q}=0)}~.
\label{22r}
\end{equation}
Here $E_b(\vec{q})$ is the free electron band energy: 
\begin{equation}
E_b(\vec{q})=-2t\sum_{i=1}^{d}\cos(q_{i}a)  
\label{23r}
\end{equation}
where $a$ is the lattice parameter and $d_{\vec{q}}(\vec{k})$ is a variational 
function that has to be determined by minimizing the expectation value 
of the Hamiltonian (\ref{1r}) on the state (\ref{20r}). 

$|\psi^{(l)}(\vec{k})>$ has the right translational symmetry, i.e. it is a 
Bloch state with wave number $\vec{k}$. This wave function represents 
an electron dressed by the virtual phonon field that describes the ionic 
polarization. We note that the term in the square brackets of the 
Eq.(\ref{21r}) 
allows a considerable advantage over the independent phonon approximation 
of Lee, Low and Pines. In the Lee, Low and Pines ansatz 
an important physical ingredient
is missing: 
it does not take into account the fact that 
the polaron energy can approach 
$\omega_0$. On the contrary the wave function (\ref{20r}) 
contains this physical 
information\cite{15}. 
In particular when the polaron excitation energy becomes equal 
to the energy of a longitudinal optical phonon, the band dispersion flattens 
and becomes horizontal. For these values of $\vec{k}$ the band has the bare 
phonon-like behaviour with very small spectral weight. For the same values 
of $\vec{k}$ and in particular at the edges of the Brillouin zone the main 
part of the spectral weight follows the bare electron band\cite{16}.  

The two wave functions (\ref{12r}) and (\ref{20r}), 
describing respectively the small and large polaron,  
differentiate mainly for the expression 
of the phonon distribution function. It is evident that, in spite of 
the assumption of no correlation (sum of coherent states in the 
phonon coordinate), the large polaron wave function takes into account 
the average effect of the correlation introduced by the electron 
recoil (Eq.\ref{22r}), effect absent in the small polaron phonon 
distribution function. 
   
The expectation value of the Hamiltonian (\ref{1r}) on 
the state (\ref{20r}) gives: 
\begin{eqnarray}
&&<\psi^{(l)}_{\vec{k}}|H_{el}|\psi^{(l)}_{\vec{k}}>=
-t\sum_{\vec{<\delta >}}e^{i\vec{k}\cdot \vec{\delta}}
e^{-\sum_{\vec{q}} |h_{\vec{q}}|^2 \left(1-e^{-i\vec{q}\cdot \vec{\delta}}
\right)} 
\left[1+\sum_{\vec{q}}\left(h_{\vec{q}}d^*_{\vec{q}}+h.c.\right)
\left(1-e^{-i\vec{q}\cdot \vec{\delta}}\right)\right. \nonumber\\
&&\left.+\sum_{\vec{q}}|d_{\vec{q}}|^2e^{-i\vec{q}\cdot \vec{\delta}}
+\sum_{\vec{q_1}}d_{\vec{q_1}}h^*_{\vec{q_1}}
\left(e^{-i\vec{q_1}\cdot \vec{\delta}}-1\right)
\sum_{\vec{q_2}}d^*_{\vec{q_2}}h_{\vec{q_2}}
\left(e^{-i\vec{q_2}\cdot \vec{\delta}}-1\right)
\right]
\label{24r}
\end{eqnarray}
\begin{eqnarray}
&&<\psi^{(l)}_{\vec{k}}|H_{ph}+H_{I}|\psi^{(l)}_{\vec{k}}>=
\left[
\sum_{\vec{q}}\left(\omega_0|h_{\vec{q}}|^2-M_qh^*_{\vec{q}}
-M^*_qh_{\vec{q}}\right)\right] 
<\psi^{(l)}_{\vec{k}}|\psi^{(l)}_{\vec{k}}>\nonumber \\
&&+\sum_{\vec{q}}\left[\omega_0|d_{\vec{q}}|^2+
\left(M_q-\omega_0h_{\vec{q}}\right)d^*_{\vec{q}}
+\left(M^*_q-\omega_0h^*_{\vec{q}}\right)d_{\vec{q}}\right]
\label{25r}
\end{eqnarray}
where 
\begin{equation}
<\psi^{(l)}_{\vec{k}}|\psi^{(l)}_{\vec{k}}>=1+\sum_{\vec{q}}
|d_{\vec{q}}|^2~.
\label{26r}
\end{equation}
Then $d_{\vec{q}}$ is fixed by the condition: 
\begin{equation}
\frac{\partial E^{(l)}_{\vec{k}}}{\partial d^*_{\vec{q}}}=0~.
\label{27r}
\end{equation}
Here $E^{(l)}_{\vec{k}}$ is the polaron energy in the weak-coupling limit: 
\begin{equation}
E^{(l)}_{\vec{k}}=\frac{<\psi^{(l)}_{\vec{k}}|H|\psi^{(l)}_{\vec{k}}>}
{<\psi^{(l)}_{\vec{k}}|\psi^{(l)}_{\vec{k}}>}~.
\label{28r}
\end{equation}
This procedure provides: 
\begin{equation}
d_{\vec{q}}=\frac{M_q-\omega_0h_{\vec{q}}
-2te^{-\sum_{\vec{q}}|h_{\vec{q}}|^2
\left(1-\cos{q_x a}\right)}h_{\vec{q}}A_{\vec{q}}(\vec{k}) }
{y_{\vec{k}}-B_{\vec{k}}+2te^{-\sum_{\vec{q}}|h_{\vec{q}}|^2
\left(1-\cos{q_x a}\right)}
\sum^{d}_{i=1} \cos{\left(k_i-q_i\right)a}}
\label{29r}
\end{equation}
where
\begin{eqnarray}
&&B_{\vec{k}}=\omega_0+\sum_{\vec{q}}\left[\omega_0|h_{\vec{q}}|^2
-M_qh^*_{\vec{q}}-M^*_qh_{\vec{q}}\right]~,\nonumber \\
&&A_{\vec{q}}(\vec{k})=\sum^{d}_{i=1}\left \{
\cos{k_ia}-\cos{\left(k_i-q_i\right)a}
+x_i(\vec{k})\left[\cos{\left(\frac{k_i}{2}-q_i\right)a}-\cos{\frac{k_i}{2}a}
\right] \right. \nonumber \\
&& \left. +z_i(\vec{k})\left[\sin{\left(\frac{k_i}{2}-q_i\right)a}
-\sin{\frac{k_i}{2}a}\right]\right \}  
\label{30r}
\end{eqnarray}
and $x_i(\vec{k})$, $z_i(\vec{k})$, $y(\vec{k})$ are variational 
parameters.

{\bf Intermediate coupling}

For any particular value of $t$ there is a 
value of the electron-phonon coupling constant ($g_{c}$) where the ground state 
energies of the two previously discussed solutions become equal.
Nevertheless the two solutions exhibit very different polaron features. 
In particular when the coupling constant is smaller than $g_c$  
the stable 
solution (the one with lowest energy) corresponds to the large polaron 
while for $g>g_c$ it corresponds to the small polaron. 
Crossing $g_c$ 
the mass of the polaronic quasi-particle increases 
in a discontinuous way\cite{17}. 
A more careful inspection shows that in this range 
of $g$ values the wave functions describing the two solutions 
of large and small polaron are not orthogonal 
and have non zero off diagonal matrix 
elements. This suggests that the lowest 
state of the system is made of a mixture of the large and small polaron 
solutions\cite{18}. 
Then the idea is to use a variational method to determine the 
ground state energy of the hamiltonian (\ref{1r}) 
by considering as trial state 
a linear superposition of the wave functions describing the two types of 
previously discussed polarons: 
\begin{equation}
|\psi_{\vec{k}}>=\frac{A_{\vec{k}} 
|\overline{\psi}^{(l)}_{\vec{k}}>+
B_{\vec{k}} |\overline{\psi}^{(s)}_{\vec{k}}>}
{\sqrt{A^2_{\vec{k}}+B^2_{\vec{k}}
+2A_{\vec{k}}B_{\vec{k}}S_{\vec{k}}}}
\label{31r}
\end{equation} 
where 
\begin{eqnarray}
&&|\overline{\psi}^{(l)}_{\vec{k}}>=
\frac{|\psi^{(l)}_{\vec{k}}>}
{\sqrt{<\psi^{(l)}_{\vec{k}}|\psi^{(l)}_{\vec{k}}>}},  
|\overline{\psi}^{(s)}_{\vec{k}}>=
\frac{|\psi^{(s)}_{\vec{k}}>}
{\sqrt{<\psi^{(s)}_{\vec{k}}|\psi^{(s)}_{\vec{k}}>}}
\label{32r}
\end{eqnarray}
and $S_{\vec{k}}$ is the overlap factor of the two wave functions 
$|\overline{\psi}^{(l)}_{\vec{k}}>$ and 
$|\overline{\psi}^{(s)}_{\vec{k}}>$: 
\begin{equation}
S_{\vec{k}}=
\frac{<\overline{\psi}^{(l)}_{\vec{k}}|\overline{\psi}^{(s)}_{\vec{k}}>+h.c.}
{2}~.
\label{33r}
\end{equation}
In the Eq.(\ref{31r}) $A_{\vec{k}}$ and $B_{\vec{k}}$ 
are two additional variational parameters 
which provide the relative weight of the large 
and small polaron solutions in the ground state of the system for any 
particular value of $\vec{k}$. 

In this paper we perform the minimization procedure in two steps. First 
the energies of the large and the small polaron wave functions are 
minimized, then these wave functions are used in the minimization 
procedure discussed in the present section. This way to proceed simplifies 
significantly the computational effort and makes all calculations 
described accessible on a personal computer. 
  
It should be noted that the trial wave function (\ref{31r}) 
contains correlation between the emission 
of successive virtual phonons in the field around the electron since 
the phonon wave function is a linear superposition of coherent states 
for any particular lattice site. Then the wave function (\ref{31r}) 
recovers, in 
the weak and strong coupling limit respectively, the large and small 
polaron wave function, introduces correlation between the emission of 
successive virtual phonons by the electron and contains the important 
physical information that the quasi-particle becomes unstable when 
the polaron excitation energy equals the energy of a longitudinal 
optical phonon.     
   
The procedure of minimization of the quantity 
$\frac{<\psi_{\vec{k}}|H|\psi_{\vec{k}}>}{<\psi_{\vec{k}}|\psi_{\vec{k}}>}$ 
with respect to $A_{\vec{k}}$ and $B_{\vec{k}}$ 
gives for the polaron energy: 
\begin{equation}
E_{\vec{k}}=\frac{E_{\vec{k}m}-S_{\vec{k}}E_{\vec{k}c}-
\sqrt{\left(E_{\vec{k}m}-S_{\vec{k}}E_{\vec{k}c}\right)^2-
\left(1-S^2_{\vec{k}}\right)
\left(E^{(l)}_{\vec{k}}E^{(s)}_{\vec{k}}-E^2_{\vec{k}c}\right)
}}
{1-S^2_{\vec{k}}}
\label{34r}
\end{equation} 
and 
\begin{equation}
\frac{A_{\vec{k}}}{B_{\vec{k}}}=
\frac{E_{\vec{k}c}-E_{\vec{k}}S_{\vec{k}}}
{E_{\vec{k}}-E^{(l)}_{\vec{k}}}~.
\label{35r}
\end{equation}
Here $E_{\vec{k}m}=\left(E^{(l)}_{\vec{k}}+E^{(s)}_{\vec{k}}\right)/2$ 
and $E_{\vec{k}c}=\left(<\overline{\psi}^{(l)}_{\vec{k}}|H|
\overline{\psi}^{(s)}_{\vec{k}}>
+h.c.\right)/2$. 
Finally the overlap factor and the matrix element of the hamiltonian 
between the two solutions $|\overline{\psi}^{(l)}_{\vec{k}}>$ and 
$|\overline{\psi}^{(s)}_{\vec{k}}>$ are, respectively: 
\begin{eqnarray}
&&<\overline{\psi}^{(l)}_{\vec{k}}|\overline{\psi}^{(s)}_{\vec{k}}>=
\sum_{\vec{R}_n}\frac{e^{i\vec{k}\cdot \vec{R}_n}}
{\left(<\psi^{(l)}_{\vec{k}}|\psi^{(l)}_{\vec{k}}>\right)^{1/2}}
\frac{\phi_{\vec{k}}(-\vec{R}_n)}
{\left(<\psi^{(s)}_{\vec{k}}|\psi^{(s)}_{\vec{k}}>\right)^{1/2}}
e^{-\sum_{\vec{q}}\left[|h_{\vec{q}}|^2+|f_{\vec{q}}|^2-2h_{\vec{q}}
f^*_{\vec{q}}e^{-i\vec{q}\cdot \vec{R}_n}
\right]/2}\nonumber \\
&&\left[1+\sum_{\vec{q}}d_{\vec{q}}\left(h^*_{\vec{q}}-f^*_{\vec{q}}
e^{-i\vec{q}\cdot \vec{R}_n}
\right)\right]
\label{36r}
\end{eqnarray}
and 
\begin{eqnarray}
&&<\overline{\psi}^{(l)}_{\vec{k}}|H_{el}|\overline{\psi}^{(s)}_{\vec{k}}>=
-t\sum_{\vec{R}_n}\frac{e^{i\vec{k}\cdot \vec{R}_n}}
{\left(<\psi^{(l)}_{\vec{k}}|\psi^{(l)}_{\vec{k}}>\right)^{1/2}}
\frac{e^{-\sum_{\vec{q}}\left[|h_{\vec{q}}|^2+|f_{\vec{q}}|^2-2h_{\vec{q}}
f^*_{\vec{q}}e^{-i\vec{q}\cdot \vec{R}_n}
\right]/2}}
{\left(<\psi^{(s)}_{\vec{k}}|\psi^{(s)}_{\vec{k}}>\right)^{1/2}}
\nonumber \\
&&\left[1+\sum_{\vec{q}}d_{\vec{q}}\left(h^*_{\vec{q}}-f^*_{\vec{q}}
e^{-i\vec{q}\cdot \vec{R}_n}\right)\right]
\sum_{<\vec{\delta}>}\phi_{\vec{k}}(-\vec{R}_n-\vec{\delta}),  
\label{37r}
\end{eqnarray}
\begin{eqnarray}
&&<\overline{\psi}^{(l)}_{\vec{k}}|H_{ph}+H_{I}
|\overline{\psi}^{(s)}_{\vec{k}}>=
\sum_{\vec{R}_n}\frac{e^{i\vec{k}\cdot \vec{R}_n}}
{\left(<\psi^{(l)}_{\vec{k}}|\psi^{(l)}_{\vec{k}}>\right)^{1/2}}
\frac{e^{-\sum_{\vec{q}}\left[|h_{\vec{q}}|^2+|f_{\vec{q}}|^2-2h_{\vec{q}}
f^*_{\vec{q}}e^{-i\vec{q}\cdot \vec{R}_n}
\right]/2}}
{\left(<\psi^{(s)}_{\vec{k}}|\psi^{(s)}_{\vec{k}}>\right)^{1/2}}
\nonumber \\
&&\phi_{\vec{k}}(-\vec{R}_n)
\left[\sum_{\vec{q}}d_{\vec{q}}\left(M^*_q-f^*_{\vec{q}}
e^{-i\vec{q}\cdot \vec{R}_n}\right) \right. \nonumber \\
&&\left. +\left(1+\sum_{\vec{q}}d_{\vec{q}}\left(h^*_{\vec{q}}-f^*_{\vec{q}}
e^{-i\vec{q}\cdot \vec{R}_n}\right)\right)
\sum_{\vec{q}}\left(
\omega_0f^*_{\vec{q}}h_{\vec{q}}e^{-i\vec{q}\cdot \vec{R}_n}
-M_qf^*_{\vec{q}}e^{-i\vec{q}\cdot \vec{R}_n}-M^*_qh_{\vec{q}}
\right)\right]~.
\label{38r}
\end{eqnarray}

\section {Numerical results}

In order to test the validity of our variational approach we recall 
the perturbative results both in the weak and strong coupling limits. 
From the weak coupling perturbative theory we get\cite{5}: 
\begin{equation}
E^{(wc)}_{\vec{k}}=E_b(\vec{k})+\Re
\left[\Sigma(\vec{k},E_b(\vec{k}))\right]
\label{5n}
\end{equation}              
where
\begin{equation}
\Sigma(\vec{k},ik_n)=\sum_{\vec{q}}\frac{|M_q|^2} 
{ik_n-\omega_0-E_b(\vec{k}+\vec{q})} 
\label{4n}
\end{equation} 
while the second order perturbation theory in the strong coupling limit 
gives\cite{6}: 
\begin{equation}
E^{(sc)}_{\vec{k}}\simeq E_p \left(1+\frac{1}{4\lambda^2}\right)
-2te^{-g^2}\cos{k_x}
-2\frac{E_p}{4\lambda^2}
e^{-g^2}\cos{2k_x} 
\label{8}
\end{equation}
in one dimension and
\begin{eqnarray}
&&E^{(sc)}_{\vec{k}}\simeq E_p \left(1+\frac{1}{8\lambda^2}\right)
-2te^{-g^2}\left(\cos{k_x}+\cos{k_y}\right)\nonumber \\
&&-2\frac{E_p}{8\lambda^2}
e^{-g^2}\left(\cos{2k_x}+\cos{2k_y}+2\cos{k_x}\cos{k_y}
\right)
\label{9}
\end{eqnarray}
in two dimensions. 

In Fig.1. and Fig.2. we report 
the polaron ground state energy obtained within our approach 
($E_{\vec{k}=0}$ in the Eq.(\ref{34r})) in one and two dimensions 
together with large and small polaron estimates (Eq.(\ref{28r}) and 
Eq.(\ref{1})),   
on which our solution is based, 
and with the perturbative results. 
As it is clear from the plots, 
our variational proposal recovers the asymptotic 
perturbative results and improves significantly both variational 
estimates in the intermediate region, where neither the perturbative 
methods nor the asymptotic variational ansatz give a satisfactory description.    
Moreover, 
our data in the intermediate region are in very good agreement with 
the results of two of the best methods available in the literature 
(see Fig.3.): 
the Global Local variational method\cite{11} and the Quantum Monte Carlo 
calculation\cite{korn}. The agreement of our results with approaches 
numerically much more sophisticated indicates that the true wave function is 
very close to a superposition of the wave functions that we have classified 
as large and small polaron solutions. The very accurate choice of the 
variational wave function has allowed a dramatic simplification of the 
numerical problem.     

Within our approach we have also studied the polaron band both in one 
and two dimensions for different values of the electron-phonon coupling 
constant (Fig.4. and Fig.5.). As for the ground state energy our 
variational ansatz is able to recover all the properties expected. 
In the weak coupling regime, increasing the value of the 
wavenumber of the polaron Bloch state, $E_{\vec{k}}$ increases until the 
excitation energy $E_{\vec{k}}-E_{\vec{k}=0}$ equals $\omega_0$. When 
$k$ is greater than this critical momentum the polaron becomes unstable 
to optical phonon emission and the dispersion curve bends over and 
becomes horizontal (this does not happen for $t/\omega_0<.25$ in one 
dimension and $t/\omega_0<.125$ in two dimensions). In the opposite regime 
the well-known polaronic band collapse takes place. Finally for 
intermediate values of the electron-phonon coupling constant the polaron 
band structure deviates significantly from both the dispersion curves. 
In particular the 
strong coupling variational result underestimates the bandwidth and 
overestimates significantly the mass enhancement. 

From our results and in agreement with Romero et al.\cite{11} 
we find that there is not qualitative difference between the polaron 
features in one and two dimensions. In both cases, also in the 
adiabatic regime, there is a range of intermediate values of the 
electron-phonon coupling constant where a crossover takes place between 
the weak coupling regime, characterized by effective masses lightly 
heavier than the free electron mass, and the strong coupling regime in 
which the well known polaronic band collapse takes place. 

Another property of interest in studying the polaronic properties  
is the ground state spectral weight: 
\begin{equation}
Z_{\vec{k}}=|<\psi_{\vec{k}}|c^{\dagger}_{k}|0>|^2
\label{6n}
\end{equation}
where $|0>$ is the electronic vacuum state containing no phonons. 
$Z_{\vec{k}}$ is the renormalization coefficient of the one-electron 
Green function and gives the fraction of the bare electron state in the 
polaronic trial wave function. In Fig.6. and Fig.7. 
we report the numerical results 
of $Z_{\vec{k}}$, at $\omega_0/t=1$, as a function of the electron-phonon 
coupling constant at $k=0$ and as a function of the polaron Bloch state 
wavenumber for different values of g. In the weak coupling regime 
$Z_{\vec{k}=0}$ is of order of the unity indicating that the polaronic 
quasi-particle is well-defined. The main part of the spectral weight is 
located at energies that correspond approximatively to the bare electronic 
levels. Instead at the edges of the Brillouin zone $Z_{\vec{k}}$ 
approaches zero. For these values of the wavenumber of the polaron Bloch 
state the main part of the spectral weight follows the bare electron band. 
Increasing the electron-phonon interaction $Z_{\vec{k}=0}$ decreases and 
approaches zero in the strong coupling regime. Here the carrier acquires 
large effective mass, the mean number of phonons in the cloud around the 
electron is very large and the most of spectral weight is located at the 
excited states, indicating that the coherent motion is suppressed 
rapidly with increasing the temperature\cite{emin}.

Finally we consider the lattice displacement associated to the 
polaron formation. An estimate of the average deviation of the 
diatomic molecule on the site $n+m$ from the equilibrium position, 
when one electron is on the site $n$, is given by the function: 
\begin{equation}
D_{\vec{k}}(\vec{R}_m)=2g\frac{S_{\vec{k}}(\vec{R}_m)}
{\sqrt{2M\omega_0}}
\label{101r}
\end{equation} 
where 
\begin{equation}
S_{\vec{k}}(\vec{R}_m)=\frac{\sum_{\vec{R}_n}
\Gamma_{\vec{k}}(\vec{R}_n,\vec{R}_m)}{2g}~. 
\label{102r}
\end{equation}
Here $M$ denotes the ionic mass and 
$\Gamma_{\vec{k}}(\vec{R}_n,\vec{R}_m)$ represents 
the correlation function between the electronic density on the site 
$n$ and the ionic displacement on the site $n+m$: 
\begin{equation}
\Gamma_{\vec{k}}(\vec{R}_n,\vec{R}_m)=
<\psi_{\vec{k}}
|c^{\dagger}_nc_n\left(a_{n+m}+a_{n+m}\right)|\psi_{\vec{k}}>~.
\label{103r}
\end{equation}  

In Fig.8. we report the numerical results of the dimensionless quantity 
$S_{\vec{k}}(\vec{R}_m)$, at $\omega_0/t=0.5$ and in one dimension, 
for different values of the electron-phonon coupling constant at 
$\vec{k}=0$. In the weak-coupling regime $S_{\vec{k}=0}(\vec{R}_m)$ 
decreases very slowly with increasing the value of $m$. This is 
consistent with the assertion that in this regime the extension of the 
polaron is large compared with the lattice parameter of the crystal. 
In the strong coupling regime $S_{\vec{k}=0}(\vec{R}_m)$ is different 
from zero only for $\vec{R}_m=0$, i.e. the lattice displacement 
is different from zero only on the cell where there is the electron, 
indicating that the quasi-particle are extremely localized. Furthermore 
the crossover from large to small polaron is very smooth.  

\section{Conclusions}

In this paper a new variational approach has been developed to investigate 
the polaron features of the Holstein molecular crystal model. 
It has been found that a simple linear superposition of Bloch states 
that describe the small and large polaron solutions provides an estimate 
of the ground state energy that is in very good agreement with the best 
results available. It has been possible to identify a range of intermediate 
values of the electron-phonon coupling constant where a crossover takes 
place between the weak and strong coupling regime. Here the small and large 
polaron wave functions are not orthogonal and both contribute to the 
formation of the so called intermediate polaron. 
We stress that the new variational approach does not require any significant 
computational effort to be implemented and 
involves, for any particular 
$k$ value, a very small number of variational parameters, that does not 
depend on the number of lattice sites. 

\section {Appendix} 
In the Holstein model ($M_q=\omega_0 g/ \sqrt{N}$)
the standard trigonometric integrals in the Eq.(\ref{1e}), 
Eq.(\ref{15r}) and Eq.(\ref{16r}) can be performed analytically. 
In one dimension 
the polaron energy in the strong coupling limit 
assumes the following form ($a=\omega_0=1$):
\begin{equation}
E^{(s)}_{\vec{k}}=\frac{
<\psi^{(s)}_{\vec{k}}|H|\psi^{(s)}_{\vec{k}}>}
{<\psi^{(s)}_{\vec{k}}|\psi^{(s)}_{\vec{k}}>}
\label{1}
\end{equation}
where
\begin{eqnarray}
&&<\psi^{(s)}_{\vec{k}}|\psi^{(s)}_{\vec{k}}>=1+
2\cos{k_x}\left(2\alpha_{\vec{k}}\beta_{\vec{k}}
+2\beta_{\vec{k}}\gamma_{\vec{k}}\right)
e^{-g^2\left(\alpha^4_{\vec{k}}+2\beta^4_{\vec{k}}+2\gamma^4_{\vec{k}}
-2\alpha^2_{\vec{k}}\beta^2_{\vec{k}}-2\beta^2_{\vec{k}}\gamma^2_{\vec{k}}
\right)}\nonumber \\
&&+2\cos{2k_x}\left(\beta^2_{\vec{k}}
+2\gamma_{\vec{k}}\alpha_{\vec{k}}\right)
e^{-g^2\left(\alpha^4_{\vec{k}}+\beta^4_{\vec{k}}+2\gamma^4_{\vec{k}}
-2\gamma^2_{\vec{k}}\alpha^2_{\vec{k}}
\right)}
+4\cos{3k_x}\beta_{\vec{k}}\gamma_{\vec{k}}
e^{-g^2\left(\alpha^4_{\vec{k}}+2\beta^4_{\vec{k}}
+2\gamma^4_{\vec{k}}-2\gamma^2_{\vec{k}}\beta^2_{\vec{k}}\right)}\nonumber \\
&&+2\cos{4k_x}\gamma^2_{\vec{k}}
e^{-g^2\left(\alpha^4_{\vec{k}}+2\beta^4_{\vec{k}}
+\gamma^4_{\vec{k}}\right)}~,
\label{2}
\end{eqnarray}
\begin{eqnarray}
&&<\psi^{(s)}_{\vec{k}}|H_{el}|\psi^{(s)}_{\vec{k}}>=
-t\left(4\alpha_{\vec{k}}\beta_{\vec{k}}+4\beta_{\vec{k}}
\gamma_{\vec{k}}\right) \nonumber\\
&&-2t\cos{k_x}\left(2\alpha_{\vec{k}}\gamma_{\vec{k}}
+\beta^2_{\vec{k}}+1\right)
e^{-g^2\left(\alpha^4_{\vec{k}}+2\beta^4_{\vec{k}}+2\gamma^4_{\vec{k}}
-2\alpha^2_{\vec{k}}\beta^2_{\vec{k}}-2\beta^2_{\vec{k}}\gamma^2_{\vec{k}}
\right)}\nonumber \\
&&-2t\cos{2k_x}\left(4\beta_{\vec{k}}\gamma_{\vec{k}}
+2\beta_{\vec{k}}\alpha_{\vec{k}}\right)
e^{-g^2\left(\alpha^4_{\vec{k}}+\beta^4_{\vec{k}}+2\gamma^4_{\vec{k}}
-2\gamma^2_{\vec{k}}\alpha^2_{\vec{k}}
\right)}\nonumber \\
&&-2t\cos{3k_x}\left(2\alpha_{\vec{k}}\gamma_{\vec{k}}
+\beta^2_{\vec{k}}+\gamma^2_{\vec{k}}\right)
e^{-g^2\left(\alpha^4_{\vec{k}}+2\beta^4_{\vec{k}}
+2\gamma^4_{\vec{k}}-2\gamma^2_{\vec{k}}\beta^2_{\vec{k}}\right)}\nonumber \\
&&-4t\cos{4k_x}\gamma_{\vec{k}}\beta_{\vec{k}}
e^{-g^2\left(\alpha^4_{\vec{k}}+2\beta^4_{\vec{k}}
+\gamma^4_{\vec{k}}\right)}
-2t\cos{5k_x}\gamma^2_{\vec{k}}
e^{-g^2\left(\alpha^4_{\vec{k}}+2\beta^4_{\vec{k}}
+2\gamma^4_{\vec{k}}\right)}
\label{3}
\end{eqnarray}
and 
\begin{eqnarray}
&&<\psi^{(s)}_{\vec{k}}|H_{ph}+H_{I}|\psi^{(s)}_{\vec{k}}>=
-g^2\left(\alpha^4_{\vec{k}}+2\beta^4_{\vec{k}}+2\gamma^4_{\vec{k}}
\right)\nonumber \\
&&+g^2\left[\left(2\alpha_{\vec{k}}\beta_{\vec{k}}
+2\beta_{\vec{k}}\gamma_{\vec{k}}\right)
\left(2\alpha^2_{\vec{k}}\beta^2_{\vec{k}}+2\beta^2_{\vec{k}}
\gamma^2_{\vec{k}}\right)
-2\alpha^3_{\vec{k}}\beta_{\vec{k}}-2\beta^3_{\vec{k}}\gamma_{\vec{k}}
-2\beta^3_{\vec{k}}\alpha_{\vec{k}}-2\gamma^3_{\vec{k}}\beta_{\vec{k}}
\right]\nonumber \\
&&2\cos{k_x}
e^{-g^2\left(\alpha^4_{\vec{k}}+2\beta^4_{\vec{k}}+2\gamma^4_{\vec{k}}
-2\alpha^2_{\vec{k}}\beta^2_{\vec{k}}-2\beta^2_{\vec{k}}\gamma^2_{\vec{k}}
\right)}\nonumber \\
&&+g^2\left[\left(\beta^2_{\vec{k}}+2\gamma_{\vec{k}}\alpha_{\vec{k}}\right)
\left(\beta^4_{\vec{k}}+2\gamma^2_{\vec{k}}\alpha^2_{\vec{k}}\right)
-2\beta^4_{\vec{k}}-2\gamma^3_{\vec{k}}\alpha_{\vec{k}}
-2\gamma_{\vec{k}}\alpha^3_{\vec{k}}
\right]2\cos{2k_x}
e^{-g^2\left(\alpha^4_{\vec{k}}+\beta^4_{\vec{k}}+2\gamma^4_{\vec{k}}
-2\gamma^2_{\vec{k}}\alpha^2_{\vec{k}}\right)}\nonumber \\
&&+g^2
\left(4\beta^3_{\vec{k}}\gamma^3_{\vec{k}}
-2\gamma_{\vec{k}}\beta^3_{\vec{k}}-2\gamma^3_{\vec{k}}\beta_{\vec{k}}\right)
2\cos{3k_x}
e^{-g^2\left(\alpha^4_{\vec{k}}+2\beta^4_{\vec{k}}
+2\gamma^4_{\vec{k}}-2\gamma^2_{\vec{k}}\beta^2_{\vec{k}}
\right)}\nonumber \\
&&+g^2\left(\gamma^6_{\vec{k}}-2\gamma^4_{\vec{k}}\right)
2\cos{4k_x}
e^{-g^2\left(\alpha^4_{\vec{k}}+2\beta^4_{\vec{k}}
+\gamma^4_{\vec{k}}
\right)}.
\label{4}
\end{eqnarray}
In two dimensions we have 
\begin{eqnarray}
&&<\psi^{(s)}_{\vec{k}}|\psi^{(s)}_{\vec{k}}>=1+
\left(\cos{k_x}+\cos{k_y}\right)\left(4\alpha_{\vec{k}}\beta_{\vec{k}}
+8\beta_{\vec{k}}\gamma_{\vec{k}}\right)
e^{-g^2\left(\alpha^4_{\vec{k}}+4\beta^4_{\vec{k}}+4\gamma^4_{\vec{k}}
-2\alpha^2_{\vec{k}}\beta^2_{\vec{k}}-4\beta^2_{\vec{k}}\gamma^2_{\vec{k}}
\right)}\nonumber \\
&&+\left(\cos{2k_x}+\cos{2k_y}\right)\left(2\beta^2_{\vec{k}}
+4\gamma^2_{\vec{k}}\right)
e^{-g^2\left(\alpha^4_{\vec{k}}+3\beta^4_{\vec{k}}+2\gamma^4_{\vec{k}}
\right)}\nonumber \\
&&+\cos{k_x}\cos{k_y}\left(8\alpha_{\vec{k}}\gamma_{\vec{k}}
+8\beta^2_{\vec{k}}\right)
e^{-g^2\left(\alpha^4_{\vec{k}}+2\beta^4_{\vec{k}}
+4\gamma^4_{\vec{k}}-2\gamma^2_{\vec{k}}\alpha^2_{\vec{k}}\right)}\nonumber \\
&&+\left(\cos{2k_x}\cos{k_y}+\cos{2k_y}\cos{k_x}\right)
8\beta_{\vec{k}}\gamma_{\vec{k}}
e^{-g^2\left(\alpha^4_{\vec{k}}+4\beta^4_{\vec{k}}
+4\gamma^4_{\vec{k}}-2\gamma^2_{\vec{k}}\beta^2_{\vec{k}}\right)}\nonumber \\
&&+\cos{2k_x}\cos{2k_y}
4\gamma^2_{\vec{k}}
e^{-g^2\left(\alpha^4_{\vec{k}}+4\beta^4_{\vec{k}}
+3\gamma^4_{\vec{k}}\right)}~,
\label{5}
\end{eqnarray}
\begin{eqnarray}
&&<\psi^{(s)}_{\vec{k}}|H_{el}|\psi^{(s)}_{\vec{k}}>=
-t\left(8\alpha_{\vec{k}}\beta_{\vec{k}}+16\beta_{\vec{k}}
\gamma_{\vec{k}}\right) \nonumber \\
&&-t\left(\cos{k_x}+\cos{k_y}\right)\left(10\beta^2_{\vec{k}}
+4\gamma^2_{\vec{k}}+2
+8\alpha_{\vec{k}}\gamma_{\vec{k}}\right)
e^{-g^2\left(\alpha^4_{\vec{k}}+4\beta^4_{\vec{k}}+4\gamma^4_{\vec{k}}
-2\alpha^2_{\vec{k}}\beta^2_{\vec{k}}-4\beta^2_{\vec{k}}\gamma^2_{\vec{k}}
\right)}\nonumber \\
&&-t\left(\cos{2k_x}+\cos{2k_y}\right)\left(4\alpha_{\vec{k}}
\beta_{\vec{k}}+16\beta_{\vec{k}}\gamma_{\vec{k}}\right)
e^{-g^2\left(\alpha^4_{\vec{k}}+3\beta^4_{\vec{k}}+2\gamma^4_{\vec{k}}
\right)}\nonumber \\
&&-t\left(\cos{3k_x}+\cos{3k_y}\right)\left(2\beta^2_{\vec{k}}
+4\gamma^2_{\vec{k}}\right)
e^{-g^2\left(\alpha^4_{\vec{k}}+4\beta^4_{\vec{k}}+4\gamma^4_{\vec{k}}
\right)}\nonumber \\
&&-t\cos{k_x}\cos{k_y}\left(16\beta_{\vec{k}}\alpha_{\vec{k}}
+48\gamma_{\vec{k}}\beta_{\vec{k}}\right)
e^{-g^2\left(\alpha^4_{\vec{k}}+2\beta^4_{\vec{k}}+4\gamma^4_{\vec{k}}
-2\alpha^2_{\vec{k}}\gamma^2_{\vec{k}}
\right)}\nonumber \\
&&-t\left(\cos{2k_x}\cos{k_y}+\cos{2k_y}\cos{k_x}\right)
\left(12\beta^2_{\vec{k}}+12\gamma^2_{\vec{k}}+8\alpha_{\vec{k}}
\gamma_{\vec{k}}\right)
e^{-g^2\left(\alpha^4_{\vec{k}}+4\beta^4_{\vec{k}}+4\gamma^4_{\vec{k}}
-2\beta^2_{\vec{k}}\gamma^2_{\vec{k}}
\right)}\nonumber \\
&&-t\cos{2k_x}\cos{2k_y} 16\beta_{\vec{k}}\gamma_{\vec{k}}
e^{-g^2\left(\alpha^4_{\vec{k}}+4\beta^4_{\vec{k}}+3\gamma^4_{\vec{k}}
\right)}\nonumber \\
&&-t\left(\cos{3k_x}\cos{k_y}+\cos{3k_y}\cos{k_x}\right)
8\beta_{\vec{k}}\gamma_{\vec{k}}
e^{-g^2\left(\alpha^4_{\vec{k}}+4\beta^4_{\vec{k}}+4\gamma^4_{\vec{k}}
\right)}\nonumber \\
&&-t\left(\cos{3k_x}\cos{2k_y}+\cos{3k_y}\cos{2k_x}\right)
4\gamma^2_{\vec{k}}
e^{-g^2\left(\alpha^4_{\vec{k}}+4\beta^4_{\vec{k}}+4\gamma^4_{\vec{k}}
\right)}
\label{6}
\end{eqnarray}
and 
\begin{eqnarray}
&&<\psi^{(s)}_{\vec{k}}|H_{ph}+H_{I}|\psi^{(s)}_{\vec{k}}>=
-g^2\left(\alpha^4_{\vec{k}}+4\beta^4_{\vec{k}}+4\gamma^4_{\vec{k}}
\right)\nonumber \\
&&+g^2\left[\left(4\alpha_{\vec{k}}\beta_{\vec{k}}
+8\beta_{\vec{k}}\gamma_{\vec{k}}\right)
\left(2\alpha^2_{\vec{k}}\beta^2_{\vec{k}}+4\beta^2_{\vec{k}}
\gamma^2_{\vec{k}}\right)
-4\alpha^3_{\vec{k}}\beta_{\vec{k}}-8\beta^3_{\vec{k}}\gamma_{\vec{k}}
-4\beta^3_{\vec{k}}\alpha_{\vec{k}}-8\gamma^3_{\vec{k}}\beta_{\vec{k}}
\right]\nonumber \\
&&\left(\cos{k_x}+\cos{k_y}\right)
e^{-g^2\left(\alpha^4_{\vec{k}}+4\beta^4_{\vec{k}}+4\gamma^4_{\vec{k}}
-2\alpha^2_{\vec{k}}\beta^2_{\vec{k}}-4\beta^2_{\vec{k}}\gamma^2_{\vec{k}}
\right)}\nonumber \\
&&+g^2\left[\left(2\beta^2_{\vec{k}}+4\gamma^2_{\vec{k}}\right)
\left(\beta^4_{\vec{k}}+2\gamma^4_{\vec{k}}\right)
-4\beta^4_{\vec{k}}-8\gamma^4_{\vec{k}}
\right]\left(\cos{2k_x}+\cos{2k_y}\right)
e^{-g^2\left(\alpha^4_{\vec{k}}+3\beta^4_{\vec{k}}+2\gamma^4_{\vec{k}}
\right)}\nonumber \\
&&+g^2\left[\left(8\alpha_{\vec{k}}\gamma_{\vec{k}}
+8\beta^2_{\vec{k}}\right)
\left(2\beta^4_{\vec{k}}+2\gamma^2_{\vec{k}}\alpha^2_{\vec{k}}\right)
-8\alpha^3_{\vec{k}}\gamma_{\vec{k}}
-16\beta^4_{\vec{k}}-8\alpha_{\vec{k}}\gamma^3_{\vec{k}}
\right]\nonumber \\
&&\cos{k_x}\cos{k_y}
e^{-g^2\left(\alpha^4_{\vec{k}}+2\beta^4_{\vec{k}}
+4\gamma^4_{\vec{k}}-2\gamma^2_{\vec{k}}\alpha^2_{\vec{k}}
\right)}\nonumber \\
&&+g^2\left(16\beta^3_{\vec{k}}\gamma^3_{\vec{k}}
-8\beta^3_{\vec{k}}\gamma_{\vec{k}}-8\beta_{\vec{k}}\gamma^3_{\vec{k}}
\right)
\left(\cos{2k_x}\cos{k_y}+\cos{2k_y}\cos{k_x}\right)
e^{-g^2\left(\alpha^4_{\vec{k}}+4\beta^4_{\vec{k}}
+4\gamma^4_{\vec{k}}-2\gamma^2_{\vec{k}}\beta^2_{\vec{k}}
\right)}\nonumber \\
&&+g^2\left[4\gamma^6_{\vec{k}}-8\gamma^4_{\vec{k}}
\right]\cos{2k_x}\cos{2k_y}
e^{-g^2\left(\alpha^4_{\vec{k}}+4\beta^4_{\vec{k}}+3\gamma^4_{\vec{k}}
\right)}.
\label{7}
\end{eqnarray}

\newpage
\begin{figure}
\noindent
\centering
\epsfxsize=0.5\linewidth
\epsffile{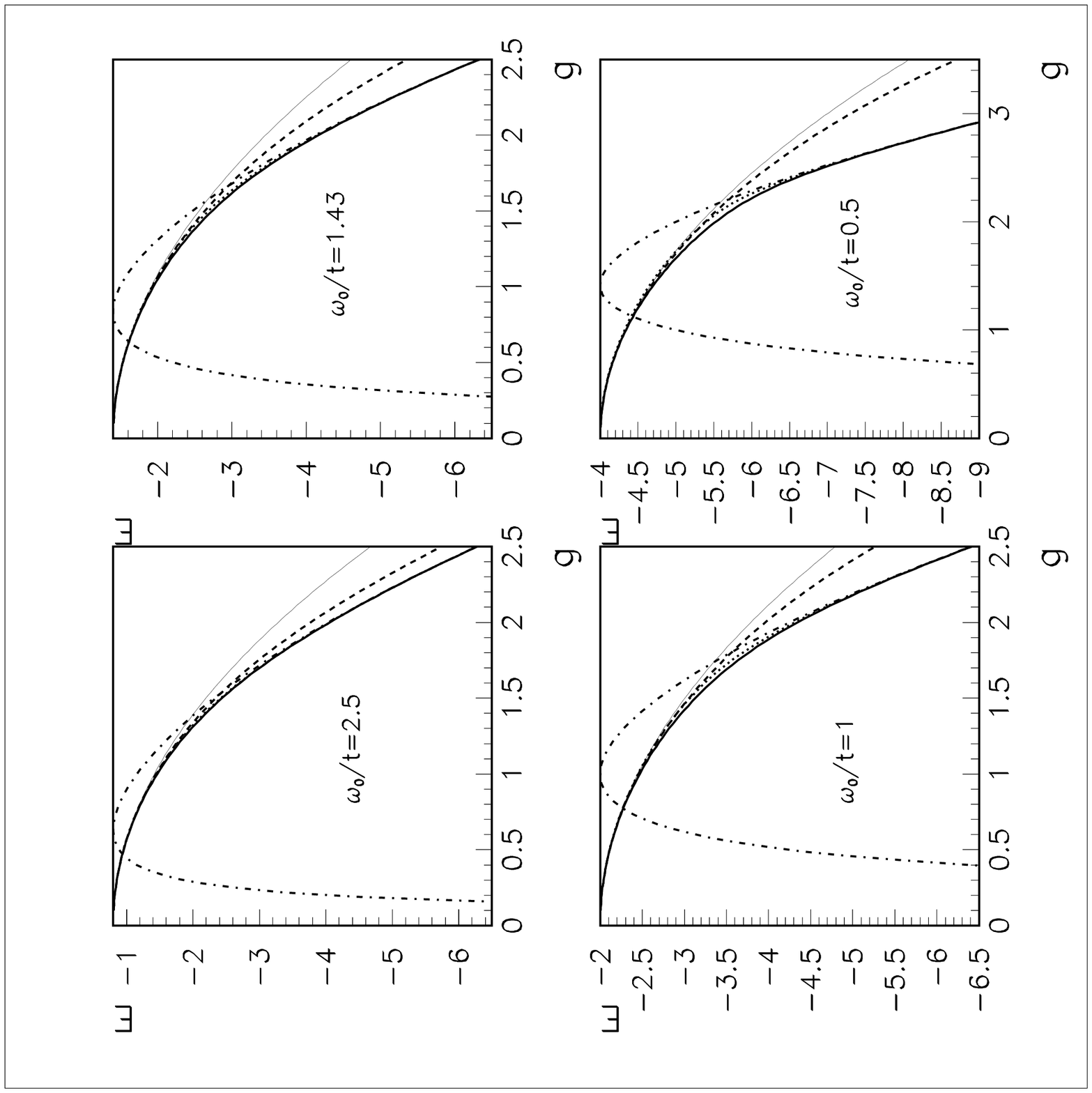}
\caption{The polaron ground state energy 
($E_{\vec{k}=0}$ in the Eq.(\ref{34r})) in one dimension (thick solid line) is
reported as a function of the electron-phonon coupling constant for different 
values of the adiabatic parameter $\omega_0 /t$. 
The data obtained within 
the approach discussed in this paper are compared with the results of 
strong $E^{(sc)}$ (Eq.(\ref{18r}), dashed-dotted line) 
and weak coupling perturbation 
theory $E^{(wc)}_{\vec{k}=0}$ 
(Eq.(\ref{5n}), thin solid line) 
and strong (dotted line) and the weak (dashed line)
coupling  
variational estimates $E^{(s)}_{\vec{k}}$ (Eq.(\ref{1})) 
and $E^{(l)}_{\vec{k}}$ (Eq.(\ref{28r})). 
The energies are given in units of $\omega_0$. }
\end{figure}
\begin{figure}
\noindent
\centering
\epsfxsize=0.5\linewidth
\epsffile{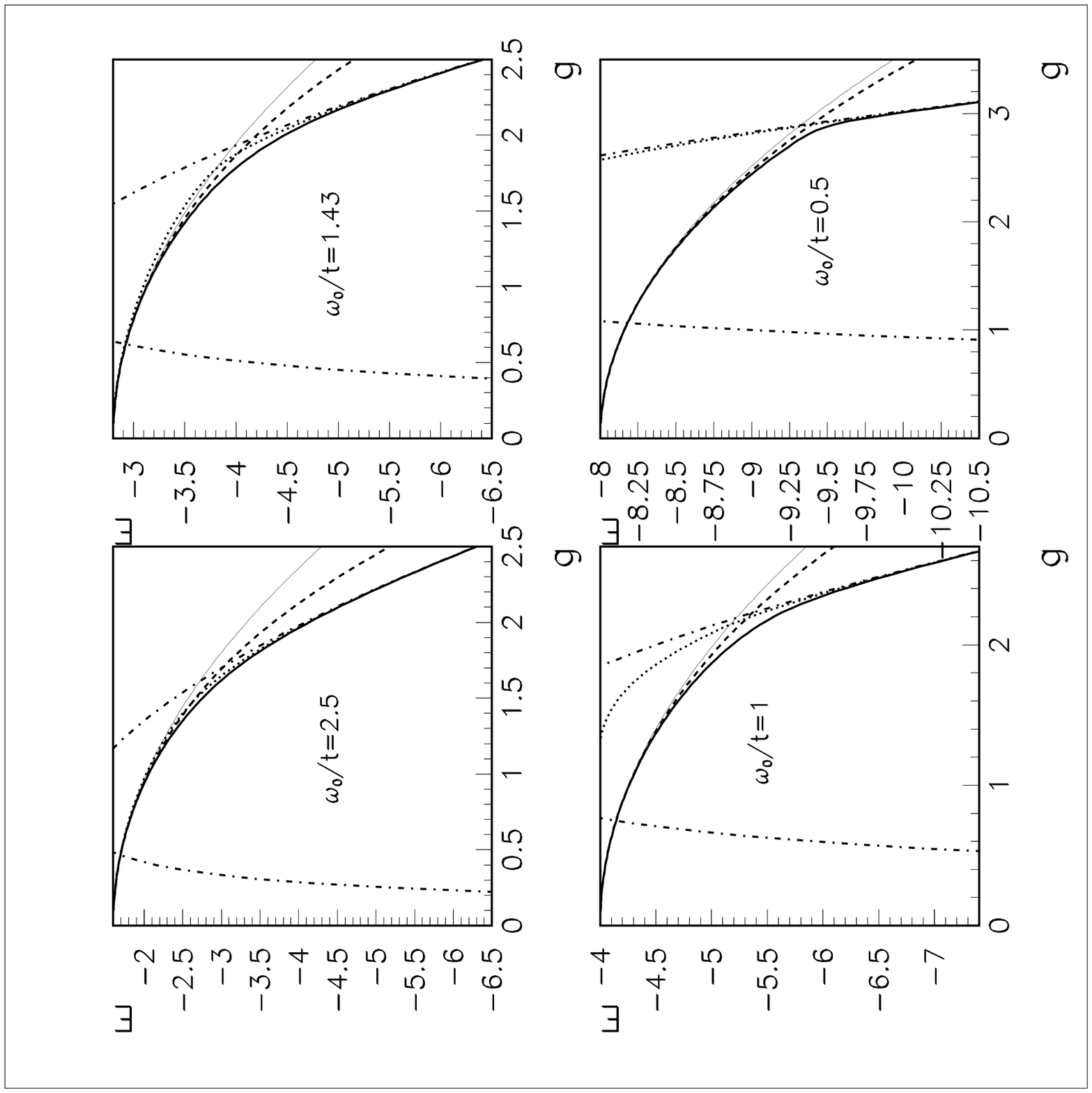}
\caption{The polaron ground state energy 
($E_{\vec{k}=0}$ in the Eq.(\ref{34r})) in two dimensions (thick solid line) 
is reported as a function of the electron-phonon coupling 
constant for different 
values of the adiabatic parameter $\omega_0 /t$. 
The data obtained within 
the approach discussed in this paper are compared with the results of 
strong $E^{(sc)}$ (Eq.(\ref{18r}), dotted-dashed line) and 
weak coupling perturbation theory $E^{(wc)}_{\vec{k}=0}$ 
(Eq.(\ref{5n}), thin solid line) 
and strong (dotted) and the weak (dashed line) coupling  
variational estimates $E^{(s)}_{\vec{k}}$ (Eq.(\ref{1}))
and $E^{(l)}_{\vec{k}}$ (Eq.(\ref{28r})). 
The energies are given in units of $\omega_0$. }
\end{figure}
\begin{figure}
\noindent
\centering
\epsfxsize=0.5\linewidth
\epsffile{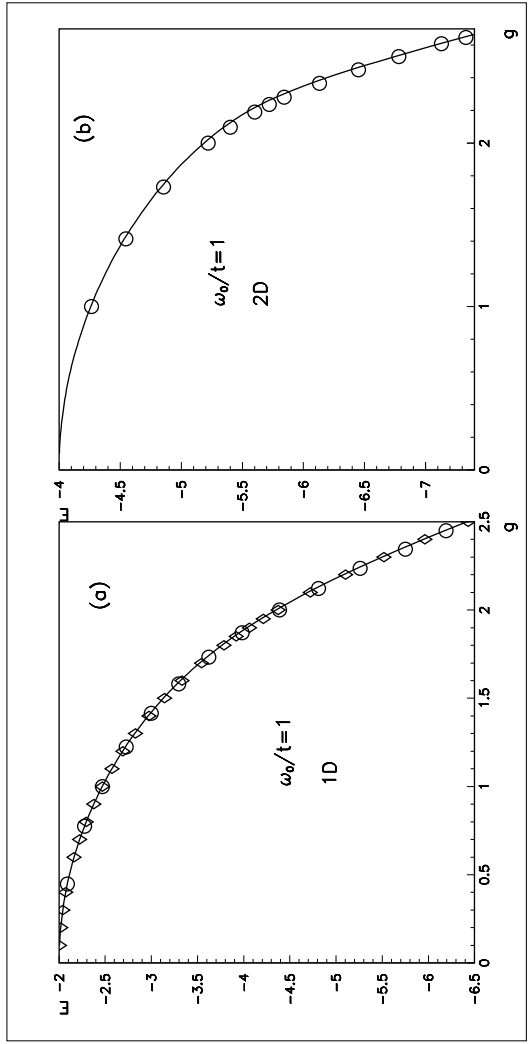}
\vspace*{1.0cm}
\caption{The variational results obtained within the approach discussed in 
this paper (Eq.(\ref{34r}), solid line)
are compared with 
the data of the Global Local variational method\cite{11} 
(diamonds), kindly provided by A. H. Romero, 
in one dimensions (Fig.3a.) and with 
the energies calculated with a Quantum Monte Carlo algorithm\cite{korn} 
(circles), kindly provided by P. E. Kornilovitch, 
in one and two dimensions (Fig.3a. and Fig.3b.) at $\omega_0/t=1$. 
The energies are given in units of $\omega_0$. }
\end{figure}
\begin{figure}
\noindent
\centering
\epsfxsize=0.5\linewidth
\epsffile{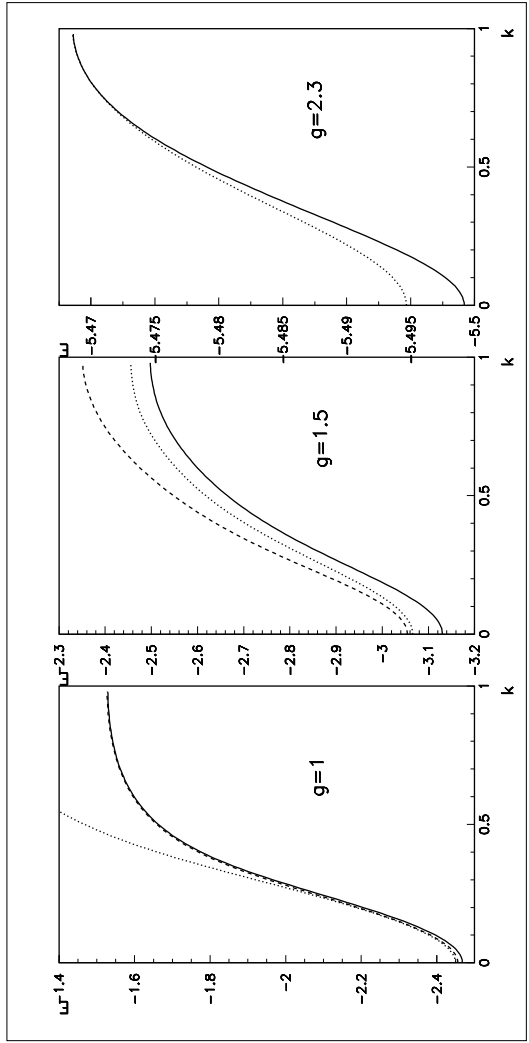}
\vspace*{1.0cm}
\caption{The polaron band structure (solid line) in one  
dimension at $\omega_0 /t=1$ is reported 
for different values of the electron-phonon 
coupling constant 
and it is compared with the weak (dashed line) and 
strong (dotted line) coupling variational estimates, 
$E^{(l)}_{\vec{k}}$ and $E^{(s)}_{\vec{k}}$. The energies and the 
momenta are given in units of $\omega_0$ and $\pi/a$ respectively.  }
\end{figure}
\newpage
\begin{figure}
\noindent
\centering
\epsfxsize=0.5\linewidth
\epsffile{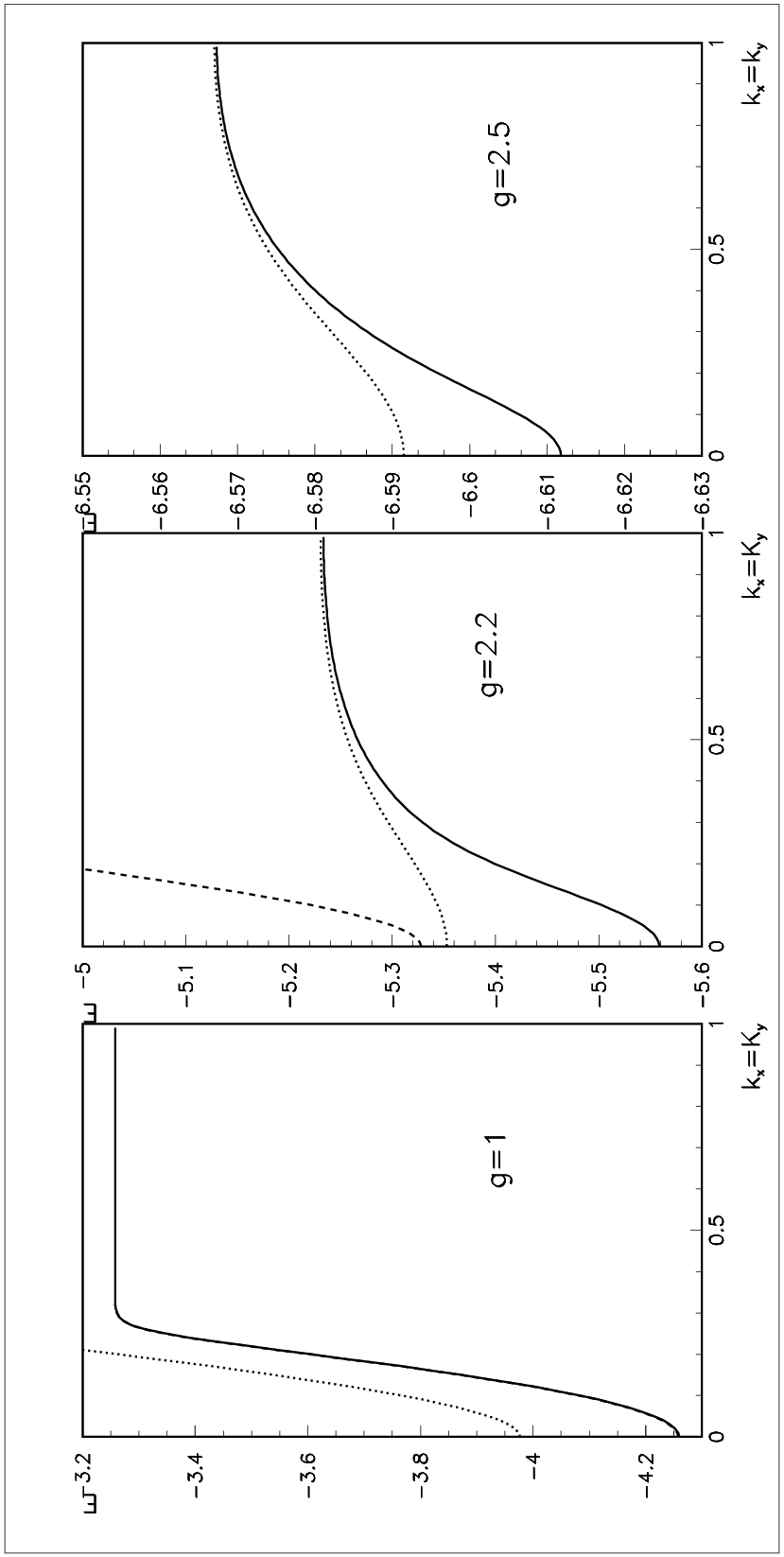}
\vspace*{1.0cm}
\caption{The polaron band structure along the diagonal ($k_x=k_y$) of the 
lattice in two 
dimensions at $\omega_0 /t=1$ is reported for different values of the 
electron-phonon 
coupling constant and it is compared with the weak (dashed line) and
strong (dotted line) coupling variational estimates,
$E^{(l)}_{\vec{k}}$ and $E^{(s)}_{\vec{k}}$. The energies and the
momenta are given in units of $\omega_0$ and $\pi/a$ respectively.   }
\end{figure}
\newpage
\begin{figure}
\noindent
\centering
\epsfxsize=0.5\linewidth
\epsffile{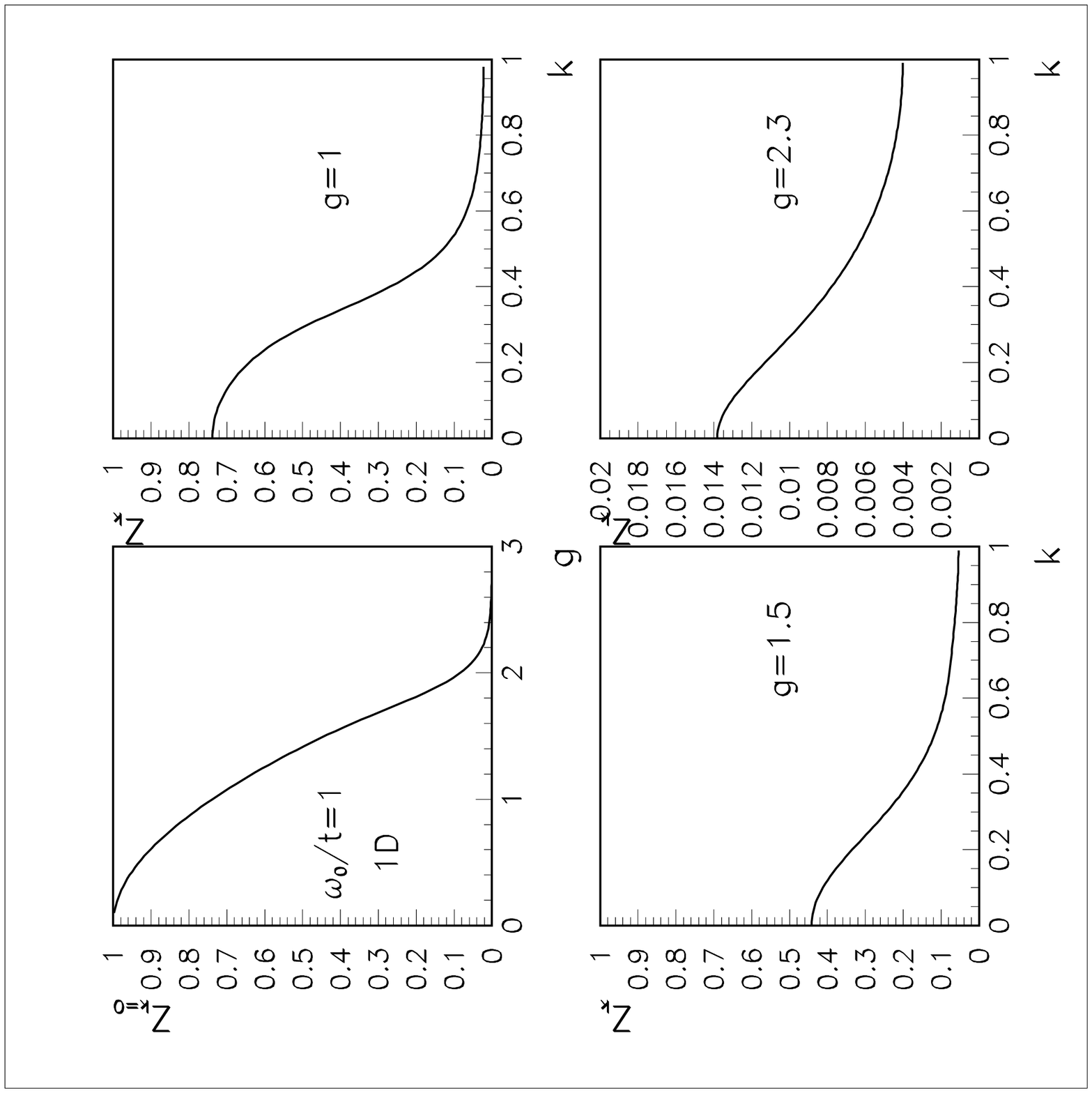}
\caption{The spectral weight of the polaronic ground state in one dimension 
as a function of the electron-phonon 
coupling constant at $k=0$ (Fig.6a.) 
and as a function of the polaron Bloch state 
wavenumber (in units of $\pi/a$) 
for different values of g (Fig.6b., Fig.6c. and Fig.6d.) 
at $\omega_0/t=1$. }
\end{figure}
\newpage
\begin{figure}
\noindent
\centering
\epsfxsize=0.5\linewidth
\epsffile{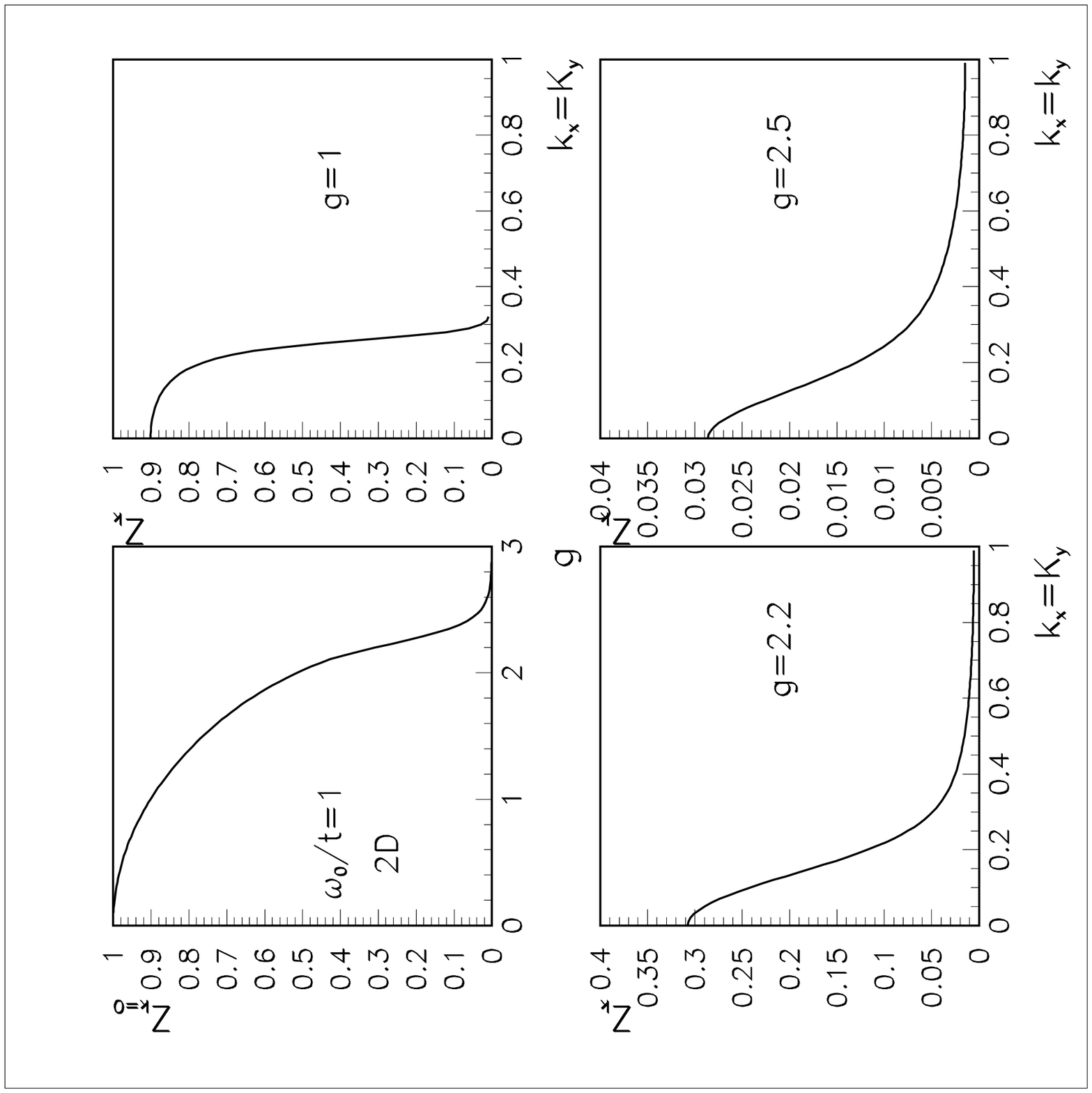}
\caption{The spectral weight of the polaronic ground state in two dimensions 
as a function of the electron-phonon 
coupling constant at $k=0$ (Fig.7a.) 
and as a function of the polaron Bloch state 
wavenumber (in units of $\pi/a$)
for different values of g (Fig.7b., Fig.7c. and Fig.7d.) 
at $\omega_0/t=1$.  }
\end{figure}
\begin{figure}
\noindent
\centering
\epsfxsize=0.5\linewidth
\epsffile{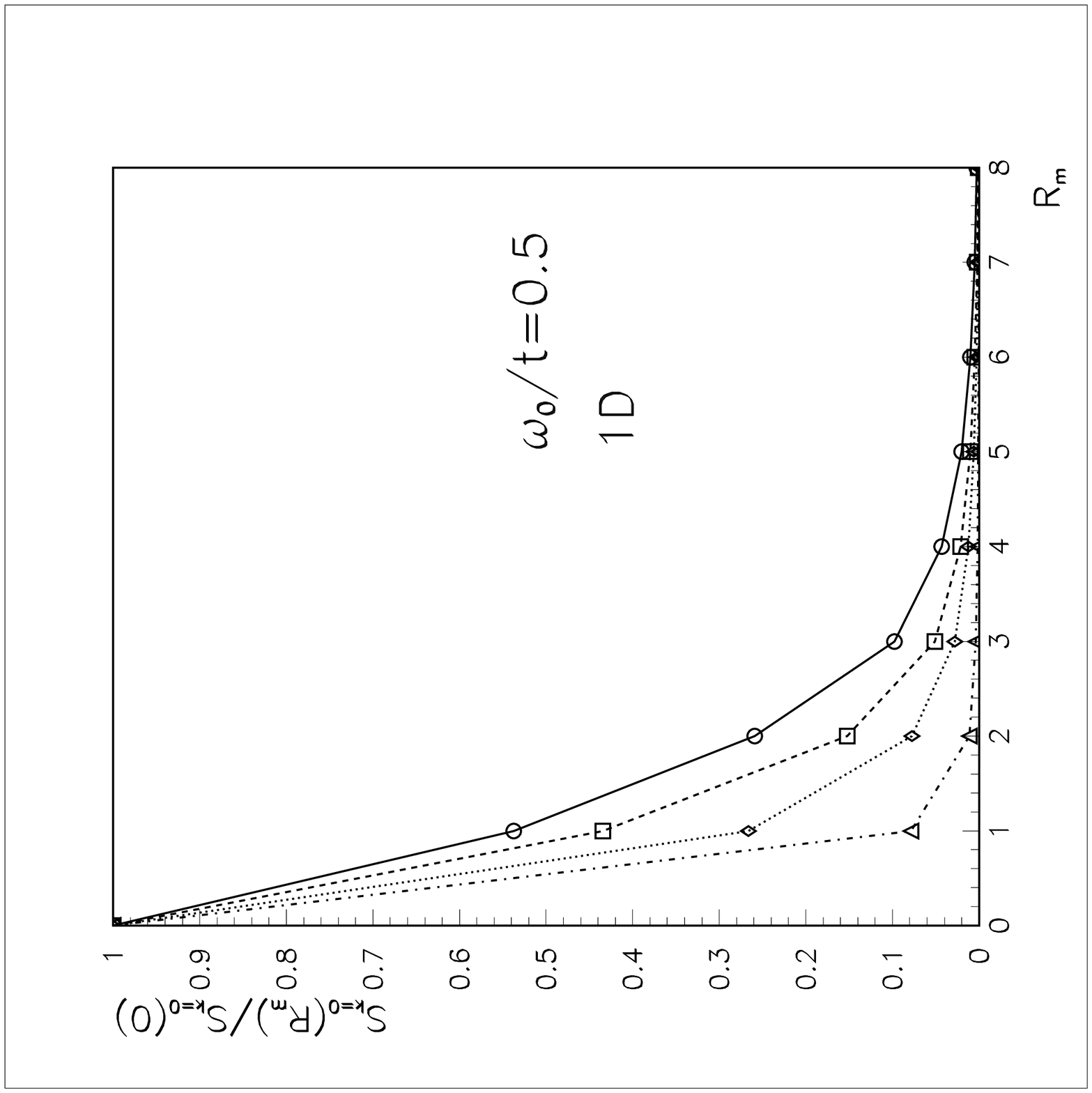}
\caption{The dimensionless quantity
$S_{\vec{k}}(\vec{R}_m)$, at $\omega_0/t=0.5$ and in
one dimension,
for different values of the electron-phonon coupling constant 
at $\vec{k}=0$: $g=1$ (circles), $g=2$ (squares), $g=2.2$ (diamonds), 
$g=2.5$ (triangles). }
\end{figure}


\begin{references} 
\bibitem {1} 
J.P. Falk, M. A. Kastner, and R. J. Birgenau,  
Phys. Rev. B {\bf 48}, 4043 (1993); 
Xiang-Xin Bi and Peter C. Eklund, Phys. Rev. Lett. {\bf 70},
2625 (1993); K. H. Kim, J. H. Jung, and T. W. Noh, 
cond-mat/9804167 (1998); 
P. Calvani, 
Proceedings of the {\it International School of Physics Enrico Fermi}, 
Course CXXXVI, Varenna (1997).  
\bibitem{2} Guo-meng-Zhao, K. Conder, H. Keller, and K. A. Muller, 
Nature {\bf381}, 676 (1996); 
Guo-Meng-Zhao, M. B. Hunt, H. Keller, and K. A. Muller, 
Nature {\bf385}, 236 (1997); 
K. H. Kim, J. Y. Gu, H. S. Choi, G. W. Park, and T. W. Noh, 
Phys. Rev. Lett. {\bf77}, 1877 (1996); 
S. J. L. Billinge, R. G. Di Francesco, G. H. Kwei, J. J. Neumeier, 
and J. D. Thompson, 
Phys. Rev. Lett. {\bf77}, 715 (1996); 
A. J. Millis, P. B. Littlewood, and B. I. Shraiman, 
Phys. Rev. Lett. {\bf 74}, 5144 (1995); 
J. M. De Teresa, M. R. Ibarra, P. A. Algarabel, 
C. Ritter, C. Marquina, J. Blasco, J. Garcia, A. del Moral and 
Z. Arnold, 
Nature {\bf 386}, 256 (1997); 
J. B. Goodenough and J. S. Zhou, Nature {\bf 386}, 229 (1997); 
J. S. Zhou, W. Archibald, and J. B. Goodenough, 
Nature {\bf 381}, 770 (1996). 
\bibitem{3} A. Lanzara, N. L. Saini, M. Brunelli, F. Natali, 
A. Bianconi, P. G. Radaelli, S. W. Cheong, 
Phys. Rev. Lett. {\bf81}, 878 (1998). 
\bibitem{Bianconi} A. Bianconi, N. L. Saini, A. Lanzara, M. Missori, 
T. Rossetti, H. Oyanagi, H. Yamaguchi, K. Oka and T. Ito, 
Phys. Rev. Lett. {\bf 76}, 3412 (1996). 
\bibitem{4} T. Holstein, Ann. Phys. {\bf8},
325 (1959)and {\bf 8}, 343 (1959);  D. Emin, Adv. Phys. {\bf 22}, 
57 (1973); for recent 
reviews on the polarons, see A. S. Alexandrov and N. F. 
Mott, Rep. Prog. Phys. {\bf57}, 1197 (1994). 
\bibitem{5} A. B. Migdal, Sov. Phys. JETP {\bf 7}, 996 (1958).
\bibitem{6} F. Marsiglio, Physica C {\bf 244}, 21 (1995); 
I. J. Lang and Yu. A. Firsov, Soviet Physics JETP {\bf16}, 1301 
(1963); Yu. A. Firsov, Polarons (Moskow, Nauka, 1975); 
A. A. Gogolin, Phys. Status Solidi {\bf109}, 95 (1982);    
W. Stephan, Phys. Rev. B {\bf 54}, 8981 (1996). 
\bibitem{7} H. de Raedt and Ad Lagendijk, 
Phys. Rev. B {\bf 27}, 6097 (1983) 
and Phys. Rev. B {\bf 30 }, 1671 (1984).  
\bibitem{korn} P. E. Kornilovitch, Phys. Rev. Lett. {\bf 81}, 
5382 (1998).  
\bibitem{8} J. Ranninger and U. Thibblin, 
Phys. Rev. B {\bf 45}, 7730 (1992); 
E. de Mello and J. Ranninger, Phys. Rev. B {\bf 55}, 14872 (1997); 
A. Kongeter and M. Wagner, J. Chem. Phys. {\bf92}, 4003 (1990);  
M. Capone, W. Stephan, and M. Grilli, 
Phys. Rev. B {\bf56}, 4484 (1997); 
A. S. Alexandrov, V. V. Kabanov and D. K. Ray, 
Phys. Rev. B {\bf49}, 9915 (1994); 
G. Wellein and H. Fehske, Phys. Rev. B {\bf 56}, 4513 (1997); 
H. Fehske, J. Loos, and G. Wellein, 
Z. Phys. B {\bf 104}, 619 (1997).
\bibitem{9} S. Ciuchi, F. de Pasquale, S. Fratini, and D. Feinberg, 
Phys. Rev. B {\bf 56}, 4494 (1997); S. Ciuchi, F. de Pasquale, 
and D. Feinberg, 
Physica C {\bf 235-240}, 2389 (1994). 
\bibitem{10} S. R. White, Phys. Rev. B {\bf 48}, 10345 (1993); 
E. Jeckelmann and S. R. White, Phys. Rev. B {\bf 57}, 6376 (1998).
\bibitem{11} A. H. Romero, D. W. Brown, and K. Lindenberg, 
Phys. Rev. B {\bf 59}, 13728 (1999); 
J. Comp. Phys. {\bf 109}, 6540 (1998); 
cond-mat/9809025 (1998) and
cond-mat/9710321 (1998). 
\bibitem{17} G. Iadonisi, V. Cataudella and D. Ninno 
Phys. Sta. Sol. (b) {\bf203}, 411 (1997); 
G. Iadonisi, V. Cataudella, G. De Filippis, and D. Ninno, 
Europhys. Lett. {\bf 41}, 309 (1998); Y. Lepine and Y. Frongillo, 
Phys. Rev. B  {\bf46}, 14510 (1992). 
\bibitem{Lowen} H. Lowen, Phys. Rev. B {\bf 37}, 8661 (1988);  
B. Gerlach and H. Lowen, Rev. Mod. Phys. {\bf 63}, 63 (1991). 
\bibitem{pekar} S. I. Pekar, Zh. Eksp. Teor. Fiz. {\bf 16}, 335 (1946); 
Ad Lagendijk and H. de Raedt, 
Phys. Lett. {\bf 108A}, 91 (1985). In these two 
references the adiabatic approximation is developed within the 
Frohlich and Holstein model respectively. 
\bibitem{13} Y. Toyozawa, Prog. Theor. Phys. {\bf26}, 29 (1961). 
\bibitem{12} H. Frohlich, H. Pelzer, and S. Zienau, 
Philos. Mag. {\bf 41}, 221 (1950); 
H. Frohlich, in {\it Polarons and Excitons}, C. G. Kuper and G. A. 
Whitfield Eds. (Oliver and Boyd, Edinburg, 1963) pag. 1. 
\bibitem{14} T.D. Lee, F. Low, and D. Pines, Phys. Rev. {\bf 90},
297 (1953). 
\bibitem{15} E. Haga, Progr. Theoret. Phys. {\bf 13}, 555 (1955); 
D. M. Larsen, Phys. Rev. {\bf 144}, 697 (1966). 
\bibitem{16} S. Engelsberg and J. R. Schrieffer, Phys. Rev. B {\bf 131}, 
993 (1963).      
\bibitem{18} D. M. Eagles, Phys. Rev. {\bf 145}, 645 (1966); 
D. M. Eagles, J. Phys. C {\bf 17}, 637 (1984); 
D. M. Eagles, Phys. Rev. {\bf 181}, 1278 (1969).
\bibitem{emin} D. Emin, Phys. Rev. B {\bf 48}, 13691 (1993). 
  
\end {references}
\end {document}